\documentclass[prb,preprint]{revtex4-1} 

\usepackage{amsmath}  
\usepackage{amsfonts} 
\usepackage{graphicx} 

\usepackage{xcolor}

\begin{document}


\title{Splitting the second: Designing a physics course with an emphasis on timescales of ultrafast phenomena}

\author{Igor P. Ivanov}
\email{ivanov@mail.sysu.edu.cn}
\affiliation{School of Physics and Astronomy, Sun Yat-sen University, 519082 Zhuhai, China}


\date{\today}

\begin{abstract}
Timescales spanning 24 orders of magnitude smaller than one second can be studied experimentally, 
and each range is packed with different physical phenomena.
This rich range of timescales offers a great context for an innovative undergraduate physics course
which introduces modern physics and technology from an unconventional perspective.
Based on the author's experience in lecturing on these topics to different audiences,
this paper proposes a syllabus of a semester-long timescale-based 
undergraduate physics course.
\end{abstract}

\maketitle 

\section{Introduction} 

The range of timescales on which physical phenomena unfold is truly staggering.
The experimentally accessible range of sub-second timescales
covers dozens of orders of magnitude,
down to about $10^{-27}$~s.
Every interval on this logarithmic scale is packed with ultrafast 
phenomena of very different nature.
Here, I advocate using this logarithmic timescale as a rich source of context
upon which an innovative, customizable lecture course can be built.
This course will introduce undergraduate science students
to the fascinating world of modern physics from an unconventional perspective.

The idea to categorize physics phenomena according to their duration is not new. 
In the academic literature, review papers on specific topics often begin 
with an overview of the spatial and temporal scales involved.
Some popular science books (Ref.~\onlinecite{tHooft:2014rjz} is a great example) 
and web pages\cite{wiki:orders-of-mag} list timescales of a large variety of phenomena.
However, these resources often contain just compilations of numbers and curious facts, 
mostly unrelated to each other.
Little effort is put into creating a coherent educational story line,
which would help understand the origins of and the interplay between timescales
in different phenomena.

A timescale-based course on ultrafast\cite{footnote-1} phenomena can overcome these shortcomings
if the lecture material is chosen and organized according to several guidelines.
First, it is a good idea to systematically cover various time intervals
starting from milliseconds and microseconds down to the shortest
measurable intervals. Within each interval, one often finds new ``protagonists'' ---
macroscopic bodies, matter, radio waves, molecules, electrons, light, subatomic particles ---
as well as characteristic analysis techniques.
Second, the course should emphasize timescale-driven connections between areas of physics
and provide concrete examples of their interplay. A few examples of these connections are given in Section~\ref{section-2}.
Third, it is instructive to explicitly demonstrate how the duration of various phenomena can be estimated
through simple order-of-magnitude calculations and to discuss limitations of these estimates.
Equipped with these examples, students can be encouraged to make numerical estimates by themselves, 
which will eventually help them to develop their own ``timescale intuition''.

The author has the experience of lecturing on these topics at different levels, 
from a single two-hour lecture for the general public \cite{my-lecture-1}
to a series of colloquia for physics researchers \cite{my-lecture-2}.
In spring 2023, the author taught this course at the School of Physics and Astronomy, 
Sun Yat-sen University (China) as an elective 
undergraduate course open to all students majoring in science.
The slides of one of these lectures can be found in the online supplementary material.
\cite{supplement}

The main goal of this paper is to inspire instructors 
to come up with their own unique timescale-based introduction
to modern physics. At the end of this paper,
a possible syllabus for a semester-long course of 36 academic hours
is outlined, but the reader should not expect to find here
a self-contained description of all the topics which could potentially be covered in this course.
The main part of the paper will, instead, 
present a small selection of illustrative situations
which exhibit timescales interplay.
Hopefully, through these examples, the reader will appreciate the variety
of timescale connections that could be shared with
undergraduate students and that could serve as an unconventional 
introduction to many branches of modern physics.

\section{Examples of timescale interplay}\label{section-2}

\subsection{Converting time into space: from oscillating droplets to high-speed cameras}

Although the millisecond range is accessible to human experience,
effects lasting only a few ms are usually too fast to be noticed by the unaided eye.
But they can be easily revealed with commercially available cameras and smartphones.
In fact, some top-level smartphones are equipped with image sensors stacked with a fast DRAM memory,
which enable the user to record short video clips at 960 fps (frames per second).
Such a smartphone, by itself, is a great tool for many in-class and at-home experiments.

How a water jet breaks up into droplets and how these water droplets oscillate are nice millisecond-range phenomena that can be studied with cameras, at home or in class. These processes are driven by capillary effects, that is the tendency of the liquid in free fall to minimize its surface area while keeping its volume unchanged. A cylindrical water column with a constant radius is unstable against certain periodic deformations of its radius because such deformations reduce the air-liquid interface area per unit length. Upon multiple pinch-offs, droplets of deformed shapes emerge and are driven by capillary forces towards the optimal shape, the sphere. As the liquid possesses inertia, droplet deformation does not stop at once when the spherical shape is attained. As a result, the droplet oscilllates between oblate and prolate shapes until the oscillations eventually die out due to viscous damping.

Although the exact mathematical description of the above processes may be extremely complicated,
their characteristic timescales can be estimated using dimensional analysis, which is a useful skill 
for students to learn.
Given the surface tension coefficient $\sigma$ of dimension $[\sigma] = \mbox{N}/\mbox{m} = \mbox{kg}/\mbox{s}^2$
and the liquid density of dimension $[\rho] = \mbox{kg}/\mbox{m}^3$,
one can construct, for a water droplet of radius $r$, the unique combination of the correct dimension:
the ``capillary time'' $\tau_c = \sqrt{\rho r^3/\sigma}$.
For water, the result is $\tau_c = 4$ ms for $r=1$ mm and 4 $\mu$s for a 10 $\mu$m droplet.
 
The capillary timescale can be used to estimate not only the droplet oscillation period
but also other capillary effects with rich dynamics 
such as a liquid bubble burst\cite{Bird:2010}, the instability and breakup of a cylindrical water jet,
with its non-trivial pinch-off and the emergence of satellite droplets.\cite{Thoroddsen:2008,Eggers:2008,VanHoeve:2010}
It may be instructive to mention that a thorough understanding of the microsecond dynamics of micrometer-sized
jets and droplets is not of pure academic interest only but is also critical for 
clean, well-controlled ink-jet printing.\cite{vanderBos:2014} 

An oscillating liquid droplet possesses another characteristic timescale, which can also be 
estimated via dimensional analysis: the viscous damping time $\tau_{visc} = \rho r^2/\eta$, 
where $\eta$ is the dynamic viscosity coefficient.
Viscous damping leads to exponential decay of the oscillation amplitude, and
if $\tau_{visc} \gg \tau_c$, which holds for mm and $\mu$m-sized water droplets, 
many oscillations take place before the oscillatory behavior dies out.
By equating the two timescales, $\tau_{visc}$ and $\tau_c$, one can determine the critical size of the droplet 
below which liquid motion is overdamped and no oscillations can occur.

How can we measure the millisecond-scale droplet oscillating period?
It can be imaged directly using a high-speed camera with thousands of frames per second (fps).\cite{Thoroddsen:2008,Eggers:2008,VanHoeve:2010}
In fact, one can find on YouTube many impressive videos of various everyday phenomena filmed 
by slow-motion enthusiasts. \cite{Youtube:slo-mo,Youtube:spaghetti}
This can also be done for large water droplets with a smartphone.
But one does not actually need to resort to the costly high-speed video recording equipment
in order to measure the droplet oscillation period.
This can be done with a single photo taken by a modest, commercially available camera.

The idea is that, when illuminated by a steady collimated light source,
a vibrating droplet reflects light in a time-dependent way.
Consider a light ray entering the droplet at an impact parameter $b$ from its center,
experiencing total internal reflection and leaving the droplet at an angle $\theta$ 
that depends on $b$. The dependence of $\theta(b)$ has an extremum $\theta_r$, 
the angle at which we observe the brightest reflection.
Since this optical phenomenon is responsible for the formation of rainbows, the angle $\theta_r$ is called the rainbow angle.
As the droplet's shape changes periodically, the optical path in a deformed droplet differs 
from that in a spherical one, which makes the rainbow angle $\theta_r(t)$ oscillate in time.
A stationary camera observing the vibrating droplet from an angle close to the rainbow angle
sees a strong periodic modulation of the reflected light intensity. 
This phenomenon has been known for more than a century and recently put to work 
in time-resolved rainbow refractometry\cite{Lv:2020}, which allows the investigation of droplet oscillations 
without imaging its shape.

\begin{figure}[h!]
	\centering
	\includegraphics[width=5in]{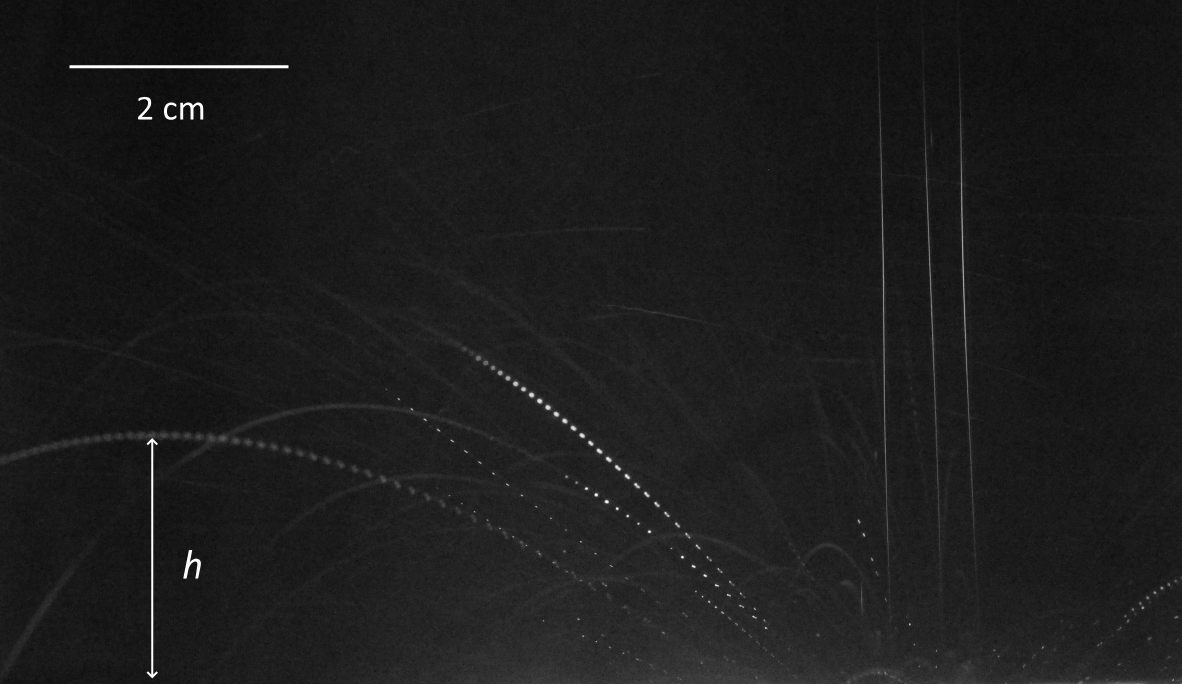}
	\caption{Measuring the water droplet oscillation period with a long exposure photo.
	The label $h$ denotes the apex height of the visible part of the parabolic trajectory.}
	\label{fig-droplet}
\end{figure}

Put simply, the vibrating droplet becomes a point-like, rapidly flickering source of light. 
If the droplet moves against a dark background, it leaves a bright dashed trajectory
on a long-exposure photograph.
By counting the number of dashes $N$ within a reference time interval $\tau$, 
one can compute the oscillation period $T = \tau/N$.
If the entire trajectory of the droplet is fully visible in the image, then $\tau$ is just the exposure time. 
But one can also rely on a different reference time defined by the free fall dynamics of the droplet itself.
Fig.~\ref{fig-droplet} shows the light traces left by water droplets falling vertically
on a hard surface (a saucer) and breaking upon impact into smaller droplets,
which fly away and, naturally, vibrate. The picture was taken with a Canon EOS 100D camera,
and, with the exposure time set to 2~s, it captured several impact instances. 
The dashed trajectories clearly show vibrations of the break-up droplets,
while the solid vertical lines, which are produced by the reflections from the large falling droplets, 
confirm that the illuminating light source was steady.
On this specific picture, one can see a parabolic trajectory 
whose apex height $h \approx 3$ cm could be determined
from the experiment geometry. The reference time $\tau$ can be calculated as the rise time: $\tau = \sqrt{2h/g} \approx 80$ ms. 
Although the air drag force affects the droplet motion, its influence is not dramatic,
which the reader can verify by observing that the trajectory does not differ too much from the symmetric parabolic curve.
Extrapolating the trajectory to the lower edge of the image frame and counting the bright dashes, 
one estimates that about 40 oscillations took place over the reference time,
which yields the oscillation period $T \approx 2$~ms.

This experiment can be the starting point to introduce a family of imaging techniques in which 
one effectively converts rapid temporal variations of light
into a spatially modulated pattern.
If the compact light source is stationary, one can take a picture
through a rotating mirror, or one can simply rotate the camera itself 
in the horizontal plane when pushing the trigger button.
With this technique, one can ``discover'' that many LED indicators found at home
are in fact blinking with frequencies up to 1 kHz and beyond.
It is even more impressive that, thanks to the persistence of human vision, 
one can train the naked eye 
to observe a similar sequence of images by swiftly moving one's gaze across 
a rapidly flickering light source.
This phenomenon known as the ``phantom array effect'' is detectable 
at flicker frequencies up to several kHz.\cite{Roberts:2012}
The author confirms observing it at least up to 1.5~kHz.

In the 19th century, the rotating or vibrating mirror technique played a role 
in studies of fast processes.
In 1855, Jules Antoine Lissajous, looking at a small light source through
a pair of perpendicularly vibrating mirrors, observed what we now call the Lissajous figures.\cite{Lissajous:1857}
In 1870s, Rudolph K\"{o}nig used the manometric flame apparatus he had invented
to visualize for the first time the acoustic waveforms produced when pronouncing various vowels.
Together with his assistant, he was observing the rapidly oscillating flame 
with the aid of a rotating mirror and sketched the patterns in his laboratory notebook.\cite{Pantalony:2004}
In the late 1850's, Berend Wilhelm Feddersen used a fast rotating mirror to measure the duration
of the spark discharge of a Leyden jar and discovered that this ultrafast process 
had a non-trivial temporal structure which included several damped oscillations
of the electric current across the spark gap.\cite{Feddersen:1858}

\begin{figure}[h!]
	\centering
	\includegraphics[width=4in]{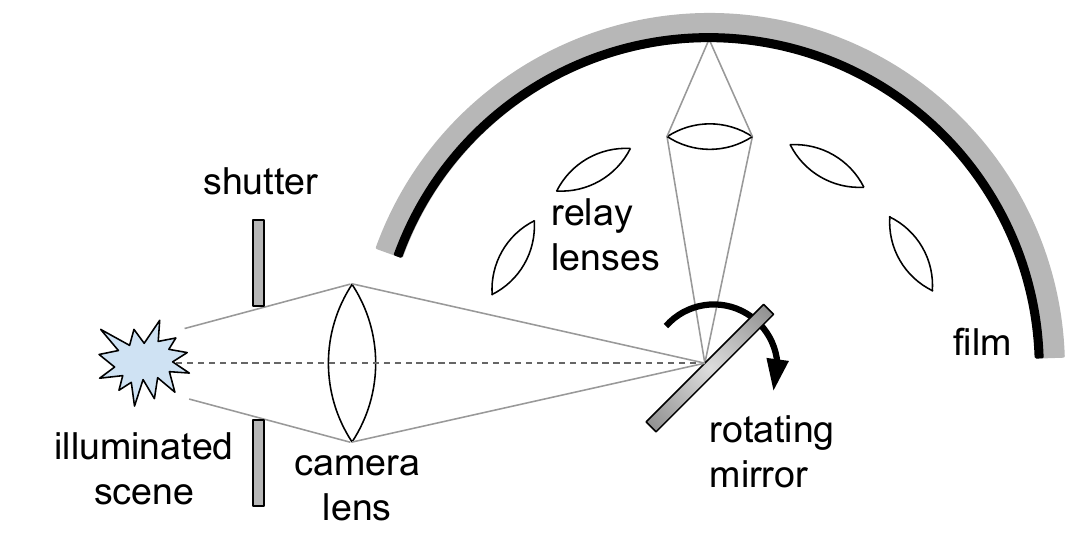}
	\caption{Principle of the rotating mirror ultrahigh-speed camera.}
	\label{fig-mirror}
\end{figure}

The rotating mirror is also a central part of a family of ultrahigh-speed cameras
capable of taking millions of frames per second (Mfps).
The high-speed cameras of the early 20th century\cite{Bull-video} reached thousands of fps,
but going beyond tens of kfps put too much mechanical stress on the film and its drive mechanism.
One of the ideas was to keep the film stationary by attaching it to the inner surface of a large hollow drum,
and to place a fast rotating mirror in the focal plane of the camera, see Fig.~\ref{fig-mirror}. 
The evolving scene of an ultrafast process was sliced by a shutter into a rapid succession of images, 
which are reflected by the rotating mirror as successive frames on the film,
allowing one to shoot short sequences at Mfps frame rates.
These cameras were initially developed during World War II to support the development of the atomic bomb,\cite{Brixner:2017} 
but their capabilities were further expanded in the post-war years.\cite{Imaging-book,Applied-imaging-book,Fuller:2009}
For instance, the CP5 camera 
had a mirror rotating at $f = 5500$ revolutions per second, which translates into the angular velocity
$\omega = 2\pi f \approx 35\,000\,\mbox{s}^{-1}$.
The angular velocity of the reflected image is twice as large, and, for the drum radius $R = 1$~m,
one gets the linear velocity of the image across the film: $v = 4\pi f R = 70\,\mbox{km/s}$.
With a 8~mm frame height plus some interframe distance, one ends up with 8~Mfps, 
which was indeed the frame rate achieved by CP5, see Chapter 25.11 of Ref.~\onlinecite{Applied-imaging-book}.
An obvious drawback of the fixed-drum rotating mirror camera is that one can only record a limited number 
of frames; CP5 had the capacity of 117 frames in one go.
Modern ultrahigh-speed cameras, such as Brandaris 128 taking up to 768 images at 25 Mfps, \cite{Brandaris}
acquire images with sensor arrays rather than films,
but they still rely on the rotating mirror technique.

The same idea --- measuring very short time intervals by converting time into space ---
is used in slit photography and in the photo finish systems 
\cite{Imaging-book,Applied-imaging-book}
and can also go beyond visible light.
The classic oscilloscope, one of the main tools of the 20th century electrical engineering, 
exploits the same principle down to the nanosecond range:
a generator sweeps the electron beam horizontally across the cathode-ray tube (CRT) screen,
and the waveform of a rapidly varying voltage, applied vertically, is seen on the screen.
Electro-optical systems, operating either in the intermittent or streak mode, 
make use of the electron beam manipulation for detecting 
rapid variations of light intensity.
The incoming light hits the photocathode producing 
a variable electron flux. The electrons are accelerated towards the CRT screen, 
but, being swept inside the tube, leave a spatially modulated image on the screen.
Coupled with a spectrograph, the electro-optical streak camera is often used in molecular physics
to obtain time-resolved spectra of rapidly evolving fluorescent light emitted
by organic molecules.\cite{Kubicki:2006}

\subsection{Ultrasound and its limits}

The shorter the timescales, the smaller are the length scales involved.
This link between the two scales becomes particularly evident
when discussing waves with an approximately linear dispersion relation, 
such as electromagnetic waves or ultrasound.

Discussing ultrasound in a undergraduate science-major class can start with the acoustic frequency scale
drawn together with the wavelength scale, with air or water as the reference medium.
After marking the audible frequency range, the instructor can ask the students 
whether the frequency scale has natural limits 
or extends infinitely in both directions.
Many students readily point out that discreteness of matter places a lower bound 
of the ultrasound wavelength $\lambda > 2a$, where $a$ is a typical intermolecular distance. 
Using the typical sound velocity in liquids and solids, $c_s\approx$ a few km/s,
one can estimate the upper end of the ultrasound frequency scale as $f = c_s/\lambda \sim 10$~THz.

Placing the lower limit on the ultrasound wavelength in air may be less intuitive for the students.
Guided by the instructor, the students should arrive at the conclusion that 
the relevant length scale is not the intermolecular distance but the much longer
mean free path in air. During this discussion, one may need to recapitulate
the concept of mean free path and emphasize its dependence on the scattering cross section.
Using the known density of air under normal conditions, the students can estimate
the mean free path ($\ell \approx 65$ nm) and find that the ultrasound frequency
scale in air ends at about 1 GHz.
Since this value depends on the air density, one can repeat the same estimate
for gases under conditions very different from normal, 
such as the extremely rarefied interplanetary medium.
Another thought-provoking question is what happens 
if a plate vibrating at a multi-GHz frequency is exposed to air.
Will it emit sound, and if not, what exactly will happen?
Such discussions will help students develop their molecular dynamics intuition.

Practical applications of ultrasound offer a rich context 
for discussing the interrelations between lengths, timescales, and frequency ranges.
The relevant literature is vast; numerous examples
can be found in historical essays \cite{Kaproth:2015,OBrien:2015},
textbooks \cite{Ultrasonics-book}, and encyclopedic resources such as 
Chapter~21 of Ref.~\onlinecite{Springer-acoustics}.
When presenting examples of ultrasound imaging, one should stress that 
the ultrasound frequency choice is often directly determined by the geometrical scale 
of the objects of interest, from bats hunting for mm-sized insects to sonography (medical ultrasound imaging) 
of various organs.
Another key parameter, which acts as a limiting factor in many applications, is 
the frequency-dependent sound attenuation in fluids. 
For example, when selecting the operating frequency in sonography, 
one seeks a balance between two requirements:
the frequency must be sufficiently high to achieve high resolution 
but, at the same time, should be low enough in order to avoid attenuation beyond the detectability limit 
of the ultrasonic signal reflected from the desired depth inside the patient's body.

Here is an example which provides some intuition on the timescales involved in sonography.
Prenatal ultrasound, a standard procedure during pregnancy,
enables observation of a moving fetus in real time. What makes this real-time video possible 
is the interplay of several timescales.
The typical operating frequency $f = 4$~MHz corresponds to an oscillation period of 0.25~$\mu$s.
A short ultrasonic pulse of duration $\tau = 1\,\mu$s is sent into the body by the transducer
and results in the depth resolution of about $c_s\tau \sim 1$~mm.
At 4~MHz, the attenuated ultrasound echo can be detected
from the depth up to about $d = 15$~cm. Once a short ultrasonic pulse is emitted, we need to wait
$\Delta t = 2d/c_s = 200\,\mu$s before sending a new pulse. 
During this ``silence period'', we record the echo returning from all the depths up to $d$.
Thus, we can obtain a one-dimensional density profile along the direction 
of the ultrasonic beam $1/\Delta t = 5000$ times per second. 
The transducer does not, of course, send the ultrasonic pulse along the same direction all the time
but rather ``rocks the beam'', that is, sweeps the directions within a certain planar angle 
within a specific plane.
Assuming that 100 one-dimensional scans along different directions 
are enough to construct a two-dimensional image, --- which is close to what modern ultrasound scanners do, 
see Chapter~21.4.3 of Ref.~\onlinecite{Springer-acoustics}, ---
we conclude that the full two-dimensional view can be updated $5000/100=50$ times per second.
This is enough for a smooth, real-time video recording.

There is also a lot of interesting physics at the high end of the frequency scale.
1~THz corresponds to an oscillation period of 1~ps
and to wavelengths of a few~nm at most.
We approach the timescales of thermal atomic or molecular motion in condensed matter
and the distances at which discreteness of matter becomes important.
As a consequence, we can expect the properties of ultrasound to change in the THz frequency range
compared with the low-frequency limit.

To illustrate this point, let us mention the curious story of the so-called ``fast sound'' in water.
In one of the first molecular dynamics (MD) simulations\cite{Rahman:1974} 
of a system of 216 water molecules performed back in 1974, 
two types of oscillation modes were observed, corresponding 
to two different sounds in water, with the high-frequency sound propagating
at $c_s \sim 3$~km/s.
The small size of the simulated system corresponded to wavelengths of 1--2~nm,
beyond the reach of the experimental methods available at the time.
A decade later, though, the high-frequency sound was confirmed experimentally,\cite{Teixeira:1985}
but its origin and properties were still debated.\cite{Ruocco:1999}
One explanation was that, at the ps scale, the ephemeral network 
of hydrogen bonds acquires extra rigidity,
which increases the speed of the usual longitudinal mode (making it the ``fast sound'').
At the same time, the ps-scale rigidity of the hydrogen bond network helps water 
elastically --- or better to say, viscoelastically --- resist to very fast shear displacements. 
Instead of the steady shear flow controlled by viscosity, 
which is observed in water at constant or slowly changing shear stress,
at ultrahigh frequencies, water supports propagating transverse vibrations, the ``transverse sound''.
In an alternative explanation, water was viewed as a mixture of two interacting fluids of heavy and light atoms,
and the fast sound was associated with certain features of the frequency vs. wavenumber curve which are typical 
for such mixtures. 

Both explanations agreed with the data at the high and at the low ends of the frequency scale, 
but differed at intermediate frequencies.
To settle the issue, it was necessary to experimentally observe how the structural modifications
of sound propagation set off in the GHz range, 
which proved to be experimentally challenging.
Indeed, MHz frequencies are well covered by traditional ultrasonic experiments, 
while the oscillation dynamics in the THz range is studied in a completely different way, 
via inelastic scattering of neutrons or X-rays. 
As a result, there existed a vast frequency gap between the two domains
exhibiting so different ultrasound properties.
Only in 2006, with the advent of new scattering experimental data
covering a significant part of the GHz range, was this gradual change of the speed of sound 
directly measured, in a clear support of the viscoelastic model.\cite{Santucci:2006}

\subsection{Positrons in matter}

Within the nanosecond range, there is a beautiful example of a technology
which stems from the interplay between the timescales of two very different effects.
The phenomenon in question is the behavior of positrons inside solid matter
before they eventually annihilate, and the technology
bears the name of the positron annihilation lifetime spectroscopy (PALS).\cite{Zaleski:2015,Gidley:2006}

A lecture on this topic can begin with a general discussion of antiparticles and antimatter.
Many students, even before taking a formal particle physics course,
are familiar with the concept of antimatter as it appears in many science fiction movies.
However their acquaintance may reveal certain misconceptions, 
such as the popular but erroneous statement that ``antiparticles travel back in time''.
After a general introduction on the topic which can straighten out the possible misconceptions,
the discussion can move on to the main question: 
what happens when a positron enters a dense medium? 
Most students know about annihilation, but they most likely have never asked themselves
how quick this process is.
They may picture it as something instantaneous:
the positron travels from vacuum into the sample
and, upon encountering the first atom, annihilates with one of its electrons.

Posed in the timescale-based course, this question by itself may
lead the students to suspect that the positron could in fact travel a certain distance 
inside the medium before annihilating.
For the quantitative discussion,
it is instructive to rephrase this question in terms of mean free path and cross section,
which are concepts students should be familiar with from the kinetic theory of gases. 
In gases, the molecules can only scatter, not annihilate,
and one calculates the mean free path from their scattering cross section.
The positron moving through a dense medium also scatters from the electrons,
$e^+e^- \to e^+e^-$, with the scattering cross section $\sigma_s$.
The direct annihilation $e^+e^- \to \gamma\gamma$ represents an additional, competing process,
characterized by its own annihilation cross section $\sigma_a$.
Thus, for an electron density $n$,
one can introduce the mean scattering length $\ell_s = 1/(n\sigma_s)$ 
as well as the mean annihilation length $\ell_a = 1/(n\sigma_a)$,
the typical distance the positron travels before annihilation.
What happens first, scattering or annihilation, 
depends on the ratio of the two cross sections.

Quantum electrodynamics allows one to accurately calculate the annihilation 
cross section.\cite{Griffiths:2008zz}
For a non-relativistic positron moving with the dimensionless velocity $\beta = v/c \ll 1$,
one gets $\sigma_a = \pi \alpha_{em}^2 \lambda_C^2/\beta$, where $\alpha_{em} \approx 1/137$
is the fine structure constant and $\lambda_C \approx 0.4$~pm is the reduced Compton wavelength of the electron.
If we assume that the positron is thermalized,
its thermal velocity being $\beta = 4\times 10^{-4}$, $v = 120$~km/s,
we can estimate the annihilation cross section: $\sigma_a = 6\times 10^{-22}$~cm$^2$.
Using the typical electron density in crystals, one obtains the annihilation length
$\ell_a \sim 100\,\mu$m, which is nearly a million times the typical interatomic distance.
For a more energetic positron with the kinetic energy in the keV range, 
which is a typical value for impinging positrons in their material science applications,
the cross section further drops by at least two orders of magnitude ($\sigma_a = 3\times 10^{-24}\,\mbox{cm}^2$ for $E_{e^+} = 1$~keV), 
and the annihilation length increases proportionally.
We arrive at the conclusion which may surprise the students:
the distance the positron travels in matter before annihilation is truly huge compared to atomic scales,
and the corresponding time is of the order of a nanosecond.

The cross section of the positron scattering in matter can also be computed with quantum electrodynamics.
Scattering of a sufficiently energetic positron from atoms is usually accompanied by atom excitation or ionization 
and leads to energy loss for the positron.
Details of the cross section calculation depend in a significant way on the energy and on the atom itself.
To get the simplest order of magnitude estimate, we can disregard the bonding energy of the atomic electrons
and view the collision as an elastic $e^+e^-$ scattering.
Furthermore, if we focus on the scattering events in which the positron direction changes by a large angle,
we can estimate the cross section as $\sigma_s \sim \alpha_{em}^2 \lambda_C^2/\beta^4$,
which, for non-relativistic positrons, is much larger than $\sigma_a$.
Alternatively, one can rely on experimental data for the positron scattering from atoms,\cite{Nahar:2020}
which show the scattering cross section in the ballpark of $\sigma_s \sim 10^{-20}\,\mbox{cm}^2$ for $E_{e^+} = 1$~keV, 
much larger than $\sigma_a$. Thus, the positron entering a dense medium will most likely experience 
multiple collisions, lose energy and, perhaps, thermalize within the timescale much shorter than a nanosecond
and well before annihilating. 

In addition to scattering and direct annihilation, the slow positron
can capture an electron to form an $e^+e^-$ bound state, the positronium.
It is known that positronium can come in two versions:
the parapositronium (p-Ps) with spin zero, which lives $\tau(\mbox{p-Ps})= 0.125$~ns
and annihilates into two photons,
and the orthopositronium (o-Ps) with spin one, which must produce
at least three photons upon annihilation and, therefore,
lives much longer: $\tau(\mbox{o-Ps})= 142$~ns.
The gap by three orders of magnitude between the two lifetimes comes from one more power 
of $\alpha_{em}$ required to create the extra photon in the o-Ps annihilation 
and a numerical coefficient
related to the phase space of the final photons
(it is, of course, up to the instructor to determine how much of this information is explained to the students).

Now comes the crucial point which enables one to measure the average pore size in a nanoporous material.\cite{Zaleski:2015,Gidley:2006}
The value $\tau(\mbox{o-Ps})= 142$~ns refers to the decay of orthopositronium in vacuum.
But inside matter, due to the constant interaction with atoms and electrons,
the lifetime can be dramatically reduced to a fraction of nanosecond.
However, the reduction is less severe if the medium is porous: 
the o-Ps has a chance to exit the dense matter 
into a pore and stick around for some time.
Measurements show that $\tau(\mbox{o-Ps})$ as a function of the pore size $d$ 
begins to depart from its vacuum value for pores smaller than 100~nm,
goes below 100~ns around $d = 10$~nm and further below 10~ns for $d\sim 1$~nm.
Thus, the o-Ps lifetime determination is a convenient method
to measure porosity and the typical pore size of nanoporous materials.
What one needs is to irradiate the sample with an intense short positron pulse,
disregard the prompt photon peak coming from parapositronium or direct $e^+e^-$ annihilation
within the first nanoseconds, and focus on the late time (up to 1~$\mu$s) 
exponential tail of the delayed photons.
In this way, one gets a nm-scale diagnostic tool
through a macroscopic counting technique.

In a highly porous material, the pores can form a network,
which the o-Ps particles can efficiently explore during their lifetime.
Taking a pore size of several nm and estimating the diffusion path length
to tens of microns, one sees that a single o-Ps can explore at least hundreds
of pores before annihilation. If sufficiently many positrons 
are implanted in the sample, some of them can meet inside a pore 
to form molecular positronium Ps$_2$; an experimental evidence of its formation
was reported\cite{Cassidy:2007} in 2007.
Let us stress again that all these remarkable phenomena are possible due 
to the significant gap between the o-Ps lifetime and the time
between its successive scattering events with the surface atoms of the medium
as it travels across the pore network.

\subsection{The puzzle of ultrafast melting}

In the previous sections, when we discussed the vibration of a water droplet or the propagation of ultrasonic waves in solids,
we were dealing with smooth motion of continuous matter. 
This was possible because the timescales involved were much longer than picoseconds, so that many molecules or atoms 
in a liquid or a solid were simultaneously affected by the phenomenon.
But around 0.1 to 1~ps, collective motion breaks down into ``elementary steps'', 
that is, jerks and collisions of individual molecules.
This molecular agitation effectively freezes as we go deep into the femtosecond range.
It is therefore natural to expect that the fastest structural changes
in condensed matter, such as ultrafast melting of crystals,
must take at least a few picoseconds.

In a typical experiment on ultrafast melting, 
one sends an intense, ultrashort focused laser pulse with duration $\ll 1$~ps 
onto a crystal sample. 
The pulse is absorbed by a thin surface layer, transmitting its energy to the electrons,
which quickly, within 100~fs, form a very hot electron gas.
However, the ions remain cold during this sub-ps absorption stage.
Indeed, they could heat up via collisions with hot electrons, 
but the electron-to-ion energy transfer is not instantaneous.
With the typical velocity of $v \sim 1000$ km/s,  
the free electrons collide with ions on the femtosecond timescale. This seems fast.
But in a single collision, the electron transmits only up to $2m_e/M_{ion} \sim 10^{-4}$
of its energy to the ion (recall the classical mechanics problem of a small mass elastically colliding 
with a massive body at rest).
Thus, one needs a few thousand electron-ion collisions to significantly heat up the ion lattice. 
This is why it takes at least a picosecond for the ultrashort laser pulse to locally melt the crystal.

The above estimate is, by itself, an instructive exercise as it illustrates the significant 
timescale gap between the dynamics of light, swift electrons and heavy, sluggish ions.
But it is also the entry point to a puzzle which took a couple of decades to solve.
Namely, starting from 1983, experimental evidence\cite{Shank:1983} kept accumulating that,
at very high pulse intensities, local melting occurs within a fraction of ps, 
significantly faster than the above estimate.
After intense studies, it was finally demonstrated\cite{Rousse:2001,Medvedev:2015} 
that this surprisingly fast melting
is of non-thermal nature. When the laser pulse removes the majority of the valence electrons from the ions,
the chemical bonding which keeps the crystal structure rigid becomes much weaker, 
and the ions readily fly away from their initial positions. 
The key point is that it happens even before the ions heat up.
Since the typical periods of the thermal vibration of atoms or ions in a crystal at room temperature 
are in the hundreds~fs range, one can expect that the crystalline order can be effectively destroyed 
within half a period or so. In a real situation, the thermal and non-thermal mechanisms compete,
but modern molecular dynamics simulations, 
which take into account the hot free electrons,
confirm that structural disorder sets in within half a picosecond after the absorption 
of an intense ultrashort laser pulse.\cite{Medvedev:2015}

\subsection{Measuring the lifetimes of unstable particles}

Metastable atomic nuclei and elementary particles have lifetimes
that span dozens of orders of magnitude,\cite{tHooft:2014rjz}
from the lifetime of the $Z$-boson\cite{Workman:2022ynf} ($\tau_Z \approx 2.6\times 10^{-25}$~s), 
the mediator of the neutral weak interaction,
to the half-lives of extremely long-lived nuclides
that exceed the age of the Universe.
Not going into the fundamental origin of lifetimes being so different,
we only mention here various methods to measure short-lived particles and nuclei.

The muon $\mu$, the heavy metastable analog of the electron, unexpectedly discovered in 1937
while studying cosmic rays,\cite{Weinberg-discovery} lives about $\tau_0(\mu) = 2\,\mu$s 
and decays to an electron and a pair of neutrinos.
Its lifetime is huge by subatomic standards; even a non-relativistic muon
can travel a macroscopic distance, which can be measured directly. 
Discussing the muon lifetime is a perfect place to introduce relativistic time dilation.
The above value $\tau_0$ refers to the muon's lifetime at rest. 
If it moves with velocity $v$, its decay, as observed in the laboratory reference
frame, slows down, and we see it live longer: $\tau = \gamma \tau_0$,
where $\gamma = (1-v^2/c^2)^{-1/2} = E/(mc^2)$ is the Lorentz factor.
In fact, without time dilation, we would not be able to detect the cosmic ray muons using ground observations.
Indeed, they are produced high in the atmosphere, at the height of about $h = 15$~km.
Dividing this height by the speed of light gives 50~$\mu$s, which is 25 times larger than $\tau_0$.
But if the initial muon energy is larger than $5$~GeV, its initial $\gamma$-factor exceeds 50.
Although such muons lose a part of their energy on their way to the ground,
most of them reach the sea level.
Numerous practical applications of muons, such as muon tomography and radiography
of cargo vehicles, large structures, and even active volcanoes,\cite{Marteau:2012zv}
would be impossible without time dilation.

The lifetimes of the so-called strange hadrons, which began appearing 
in accelerator experiments in the late 1940s,\cite{tHooft:2014rjz,Weinberg-discovery}
cluster around 0.1--1~ns, with the exception of kaons which live tens of nanoseconds.
But a nanosecond multiplied by the speed of light gives
the macroscopic distance of 30~cm, which grows further when time dilation is accounted for.
Thus, exploration of strange hadrons poses no problem: one can directly observe 
the ionization tracks left by the charged strange hadrons in a sensitive medium.\cite{Weinberg-discovery}

The collider experiments of the 1970s and 1980s brought to light two new families of particles:
the charmed and beauty hadrons.\cite{Griffiths:2008zz} These hadrons contain a heavy quark, $c$ or $b$,
which decays via weak interactions. Interestingly, these two families have lifetimes
that fill a rather narrow interval: from about 0.1~ps to 2~ps.\cite{tHooft:2014rjz,Workman:2022ynf} 
Three orders of magnitude shorter than the strange hadron decays, 
these lifetimes can still be measured directly by detecting the distance these particles travel 
between the production and decay points.

\begin{figure}[h!]
	\centering
	\includegraphics[width=4in]{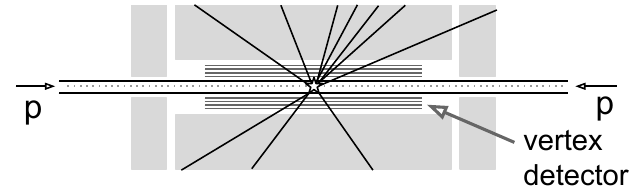}
	\bigskip

	\hspace{1.8cm}\includegraphics[width=4.5in]{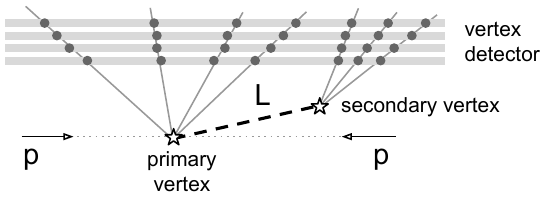}
	\caption{Top: schematic side view of a cylindrical particle detector at a proton collider. 
		Charged particles produced in proton (p) collisions leave tracks across the detector, 
		which are measured with high accuracy in the vertex detector.
		Bottom: zoomed view on the central part of the detector, showing the first few layers 
		of the vertex detector.
		A high-energy proton collision takes place at the primary vertex and leads to the production
		of many hadrons, which can include a short-lived hadron (heavy dashed track). 
		The short-lived hadron travels distance $d$ inside the beam pipe and decays into daughter hadrons 
	at another point (the secondary vertex).
	By measuring and extrapolating the trajectories of all the charged particles,
	the vertex detector accurately locates the primary and secondary vertices and 
	measures their sub-mm distance.}
	\label{fig-vertices}
\end{figure}

To illustrate how this is done in modern collider experiments, imagine a high-energy collision 
taking place inside a vacuum pipe with a diameter of a few centimeters.
Tens of charged particles can be produced in this collision.
If long-lived, they exit the beam pipe and pass through 
the sensitive elements of a multi-layered cylindrical detector.
The tracks they leave across the detector seem to emerge from a single primary vertex, 
the spot where the collision took place, see Fig.~\ref{fig-vertices}, top.

Detecting a short-lived hadron containing a heavy quark (let it be the $D$-meson, for definiteness)
and measuring its lifetime in such a busy environment are not easy tasks.
One picosecond multiplied by the speed of light gives the distance $d = 0.3$~mm. 
Both the production point (the primary vertex) and the decay point (the secondary vertex) of the $D$-meson 
are located inside the vacuum pipe.
As a result, the detector sees not the $D$-meson itself but only its decay products. 

Fortunately, the very central part of the detector, called the vertex detector, 
can accurately reconstruct the passage of individual particles through each of its cylindrical layers, 
see Fig.~\ref{fig-vertices}, bottom.
Extrapolating each trajectory, we can find the point inside the beam pipe 
where it crosses with the trajectories of many other particles.
In this way, we can distinguish hadrons emitted from the primary vertex 
and a group of particles emerging from the secondary vertex lying at a distance.
The location of the two points can be determined with an accuracy of tens of micrometers,
which is enough to measure picosecond-scale lifetimes.

\begin{figure}[h!]
	\centering
	\includegraphics[width=4in]{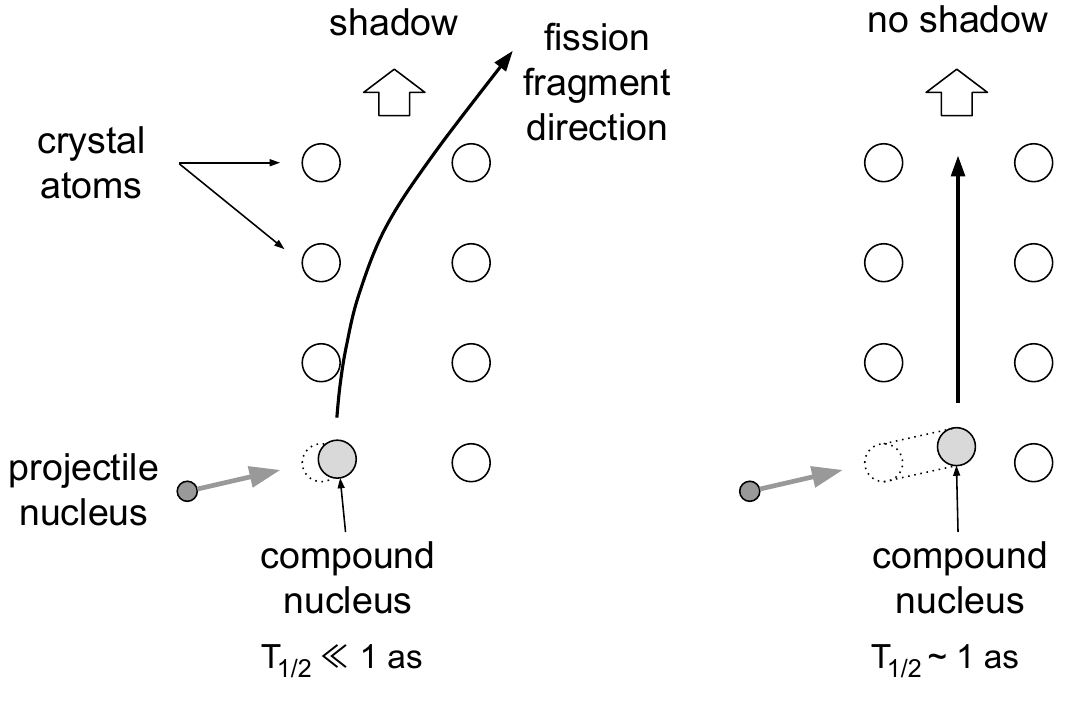}
	\caption{The principle of the crystal blocking technique to detect
	the attosecond-scale half-life of an unstable nucleus.}
	\label{fig-crystal}
\end{figure}

Actually, the displacement of an unstable particle can be detected even if it does not exceed the atomic scale.
This method is called the crystal blocking technique and is used in nuclear physics
to detect unstable nuclei with half-lives of the order of attosecond ($1\,\mbox{as}=10^{-18}\,\mbox{s}$).
Consider a projectile nucleus traveling at speed $\sim 0.2\,c$ and colliding
with a target nucleus in a crystal to produce a short-lived heavy isotope, 
Fig.~\ref{fig-crystal}.
Due to momentum conservation, this heavy short-lived nucleus keeps moving, 
its velocity being of the order of $\sim 0.1\,c$.
However, if the compound nucleus half-life is much shorter than 1~as,
it shifts from its original position by a distance much smaller than $1\,\mbox{as}\times 0.1\,c  = 0.3\,\mbox{\AA}$, 
and its fission occurs within the same crystallographic plane.
The fission products are stable, can be emitted at different angles and leave the sample
approximately keeping its initial direction. 
But those fission fragments which are initially emitted along the plane will be deflected away
by the repulsive Coulomb force of the ions sitting within the same crystallographic plane.
Therefore, a detector measuring the angular distribution of the fission fragments 
will observe depletion (the ``shadow'') of fragments in the direction parallel to the crystallographic plane,
Fig.~\ref{fig-crystal}, left.
On the other hand, if the nucleus survives for at least 1~as, 
it has time to move into the space between the planes, Fig.~\ref{fig-crystal}, right.
The fission products emitted along the crystallographic plane will not be significantly deflected,
since their trajectory is sufficiently far away from the other nuclei in the same plane.
The detector will now see fission fragments emitted in all directions, including parallel to the plane,
and the shadow gets weaker or even disappears altogether.
This method was used\cite{Morjean:2008}  to detect the as-scale lifetime of 
certain isotopes of elements 120 and 124.

Shorter lifetimes, belonging to the
zeptosecond ($1\, \mbox{zs} = 10^{-21}$~s) and yoctosecond ($1\, \mbox{ys} = 10^{-24}$~s) time ranges,
are measured indirectly, via the energy-time uncertainty relation.\cite{Busch:2008}
Applied to an unstable particle, it links its lifetime $\tau$ 
with the width $\Gamma$ of its resonance curve on the plot 
of the relevant cross section as a function of the collision energy:
$\tau = \hbar/\Gamma$, where $\hbar$ is the reduced Planck's constant
which can be conveniently expressed as 
$\hbar \approx 0.66 \,\mbox{fs}\cdot\mbox{eV} =  0.66 \,\mbox{zs}\cdot\mbox{MeV}$. 
For example, in a very recent study,\cite{Duer:2022ehf} the tetraneutron ${}^4 \mbox{n}$ --- a quasibound state
of four neutrons without any proton, --- was unambiguously observed
in the transfer reaction $\mbox{p} + {}^8\mbox{He} \to \mbox{p} + {}^4 \mbox{He} + {}^4 \mbox{n}$
as a clear peak when plotting the number of detected events as a function of the energy of the four-neutron system.
Its width $\Gamma \approx 1.75$ MeV corresponds, 
via the uncertainty relation, to the tetraneutron lifetime of $0.4$~zs,
one of the shortest timescales ever measured in nuclear physics.
 
In the world of elementary particles, excited hadrons appear as broad resonances 
in the hadron scattering cross sections with $\Gamma \sim 100$ MeV, 
which translates into 7~ys. This is close to the ``hadronic time'' $\tau_h$, the minimal timescale
on which we can talk about hadrons. It can be estimated by 
dividing the hadron size $\approx 1$~fm by the speed of light, which gives $\tau_h \approx 3$~ys.
In what concerns particle lifetimes, the record holder is the $Z$-boson,
the heavy mediator of the neutral weak interactions. 
It is visible as a spectacular resonance peak in the cross section of the $e^+e^-$ annihilation into hadrons,
standing nearly a thousand times higher than the non-resonant cross section, see Chapter 53 of Ref.~\onlinecite{Workman:2022ynf}.
Measured at the LEP collider at CERN with unprecedented precision, it exhibits the width $\Gamma_Z \approx 2.5$ GeV, 
from which one deduces the extremely short lifetime $\tau = 0.26$~ys
(or 260~rontoseconds,
if we use the recently approved\cite{Gibney:2022} SI prefix ronto-, $10^{-27}$). 

\section{Building the syllabus}

The examples described above show what kind
of timescale-based lessons could be drawn
from modern physics and technology.
However, bringing them together in a single coherent course requires
substantial work from the instructor, both in terms of learning 
and selecting the material.  

Table~\ref{table-plan} outlines a possible syllabus 
for a semester-long undergraduate course of 36 academic hours.
In addition, the online supplementary material\cite{supplement} contains 
the slides of one of the author's lectures within the course 
``The world of ultrafast phenomena'' given at the Sun Yat-sen University, China. 

The possible syllabus in Table~\ref{table-plan} covers a large variety of areas, spanning from classical physics
and development of measurement technology 
to quantum topics including nuclear and particle physics.
This collection of topics is by no means complete nor obligatory;
each instructor can build a unique course according 
to their own preferences.
Since the course is aimed at undergraduate physics
and science-major students who may not have had a quantum mechanics course, 
the advanced physics topics are not meant to be taught systematically.
The instructor is supposed to select vivid timescale-related illustrations, 
which could be presented at that level together with basic numerical estimates,
and put them in appropriate context.

About one third of the syllabus proposed in Table~\ref{table-plan} 
deals with the history of various branches of physics and technology. 
Experience shows that, with such historical excursions, the audience
appreciates the human side of the physical world exploration
and gets a better perspective on today's achievements and challenges.
Certainly, this material can be shortened or expanded 
according to the instructor's plan.
For example, a 18-hour syllabus could skip lengthy historical excursions
and reduce the number of illustrative examples,
keeping the major areas mentioned in the table
and mentioning the main experimental techniques such
as direct imaging, ultrashort laser pulses and the pump-probe method, 
and spectral analysis in various forms including 
the energy-time uncertainty relation.

There are other degrees of freedom along which the course could be customized,
such as the level of the target audience.
When lecturing to physics graduate students and researchers, 
one can assume a higher level of physics knowledge including some quantum topics.\cite{my-lecture-2}
The syllabus could include more practical applications, with further technical details,
and interesting connections across physics domains rooted in (mis)matching timescales.
In contrast, a series of popular science lectures to high-school students
may skip many technical and historical aspects 
and focus on visual content, on interactive discussions which stimulate imagination, 
on experimental demonstrations and educational games. 
But even at this level, the lectures should actually teach the school students a few basic skills
such as making numerical estimates based, at least, on the uniform motion formula.

It is also possible to deliver a single two-hour popular science lecture
in which one can introduce the entire range of experimentally accessible timescales 
and illustrate it with a selection of remarkable phenomena.\cite{my-lecture-1}  
If needed, such a lecture could also include geophysical and astronomical phenomena of very long duration, 
see Ref.~\onlinecite{tHooft:2014rjz} for possible topics and examples.

In summary, there is immense amount of educational material related 
to durations of physical processes, which is not yet fully exploited beyond pure academic publications.
Timescales much shorter than one second can become a central theme 
in a unique undergraduate physics course which 
would introduce modern physics from an unconventional perspective.
The length, the selection of material, the depth of treatment, the amount of historical details
can be adjusted according to the instructor's preferences.
As there is no single textbook which could provide 
all the necessary material,
the motivated instructor should invest significant time and effort into learning a vast range of topics
and building his/her own syllabus.
But the outcome will be rewarding. In this paper I tried to illustrate
what kind of non-trivial interplay between timescales of different phenomena
could be shown in this course.
I hope that these examples and the possible syllabus outlined in Table~\ref{table-plan}
will inspire physics teachers and educators to build their own unique timescale-focused lecture courses.

\section*{Author declarations}

The author has no conflicts to disclose.

\begin{table}[h!]
\centering
\caption{A possible syllabus for a 36-hour course}
\begin{ruledtabular}
\begin{tabular}{p{10cm} c c}
topic & timescale ranges & hours \\ \hline	
Everyday phenomena, fracture, droplets and jets, photographic and smartphone experiments & ms, $\mu$s & 4 hrs \\
High-speed imaging: 19th century photography (exposure time from minutes to microseconds), 
evolution of ultrahigh-speed cameras, the current state-of-the-art & ms, $\mu$s, ns, ps & 2 hrs \\
Acoustics and ultrasonics: history, experiments, basic spectral analysis, 
ultrasound applications & ms, $\mu$s, ns, ps & 4 hrs \\
Electric oscillations and electromagnetic waves: 
history (from Leyden jar to electric circuits), cathode ray tube and oscilloscope,
discovery of radio waves, AM/FM modulation
& $\mu$s, ns, ps & 4 hrs \\
From matter to atoms: collective motion at the ns timescale
(crack propagation, ablation, sonoluminescence),
molecular motion and molecular dynamics simulation,
protein folding, extreme ultrasound & ns, ps & 4 hrs \\
The atom: the atomic size, Bohr's model, timescale estimates, excited atoms and their lifetimes & down to as  & 2 hrs \\
The femtosecond world: molecular oscillations, electrons in atoms, femtochemistry and femtobiology & fs, as & 2 hrs \\
Short laser pulses: history, key parameters, pump-probe technique, attosecond science & ns, ps, fs, as & 4 hrs \\
Nuclear processes: estimates, nuclear processes and their timescales, heavy helium isotopes,
uranium isotopes, time-energy uncertainty relation, experimental techniques
& down to zs & 4 hrs \\
Elementary particles: estimates, lifetimes and time dilation, 
experimental techniques, relativistic nuclei collisions
 & down to ys and below & 4 hrs \\
Students' presentations & --- & 2 hrs \\ 
\end{tabular}
\end{ruledtabular}
\label{table-plan}
\end{table}


\begin{thebibliography}{99}

\bibitem{tHooft:2014rjz}
G.~'t Hooft and S.~Vandoren,
\textit{Time in powers of ten: Natural phenomena and their timescales}
(World Scientific, 2014).

\bibitem{wiki:orders-of-mag}
Wikipedia page ``Orders of magnitude (time)'',
\url{<https://en.wikipedia.org/wiki/Orders_of_magnitude_(time)>}.

\bibitem{footnote-1}
I focus here on ultrashort timescales because the world of ultrafast phenomena tends to be richer than of ultraslow processes. The proposed course could also include topics from geophysics and astronomy which involve timescales much longer than one second.

\bibitem{my-lecture-1}
Igor Ivanov, ``Splitting the second,'' popular science lecture (in Russian),
\url{<https://www.youtube.com/live/pWSWjaPGPiM?si=A-SYQjRE1_02oy9S>}

\bibitem{my-lecture-2}
Igor Ivanov, ``Timescales: Travelling Deep into the Second'', a series of educational lectures 
given at JINR, Dubna (Russia),
\url{<https://dlnp-update.jinr.ru/en/education-outreach/lectures/1029>}

\bibitem{supplement}
Igor Ivanov, ``Acoustics and ultrasound'', lecture given at the Sun Yat-sen University, China.
A URL to the supplementary material will be added by AIP.


\bibitem{Bird:2010}
James C. Bird et al. ``Daughter bubble cascades produced by folding of ruptured thin films,'' 
Nature \textbf{465} (7299), 759--762 (2010).

\bibitem{Eggers:2008}
Jens Eggers, Emmanuel Villermaux, ``Physics of liquid jets,'' 
Rep. Prog. Phys. \textbf{71} (3), 036601 (2008).

\bibitem{VanHoeve:2010}
Wim van Hoeve, Stephan Gekle, Jacco H. Snoeijer, Michel Versluis, Michael P. Brenner, and Detlef Lohse, 
``Breakup of diminutive Rayleigh jets,'' Phys. Fluids \textbf{22} (12), 122003 (2010).

\bibitem{Thoroddsen:2008}
Sigurdur T. Thoroddsen, Takeharu Goji Etoh, and Kohsei Takehara, 
``High-speed imaging of drops and bubbles,'' 
Ann. Rev. Fluid Mech. \textbf{40}, 257--285 (2008).

\bibitem{vanderBos:2014}
Arjan van der Bos et al, 
``Velocity profile inside piezoacoustic inkjet droplets in flight: comparison between experiment and numerical simulation,'' 
Phys. Rev. Applied \textbf{1} (1), 014004 (2014).

\bibitem{Youtube:slo-mo}
Youtube channel ``The Slow Mo Guys'',
\url{<https://www.youtube.com/user/theslowmoguys>}.

\bibitem{Youtube:spaghetti}
``Secret of snapping spaghetti in slow motion'', 
Youtube video on channel Smarter Every Day,
\url{<https://youtu.be/ADD7QlQoFFI>}.

\bibitem{Lv:2020}
Q. Lv, Y. Wu, C. Li, X. Wu, L. Chen, K. Cen, 
``Surface tension and viscosity measurement of oscillating droplet using rainbow refractometry,'' 
Opt. Lett. \textbf{45}, 6687--6690 (2020).

\bibitem{Roberts:2012}
J.~E.~Roberts and A. J. Wilkins, ``Flicker can be perceived during saccades at frequencies in excess of 1 kHz,'' 
Lighting Research \& Technology \textbf{45} (1), 124--132 (2013).

\bibitem{Lissajous:1857}
J.~Lissajous, ``Mémoire sur l'étude optique des mouvements vibratoires,''
Ann. Chim. Phys. \textbf{51}, 147 (1857).

\bibitem{Pantalony:2004}
David Pantalony, 
``Seeing a voice: Rudolph Koenig's instruments for studying vowel sounds,'' 
Am. J. Psychology \textbf{117} (3), 425--442 (2004).

\bibitem{Feddersen:1858}
B. W. Feddersen, ``LVIII. Contributions to the knowledge of the electric spark,'' 
The London, Edinburgh, and Dublin Philosophical Magazine and Journal of Science 
\textbf{16} (110), 503--516 (1858).


\bibitem{Bull-video}
Youtube video of Lucien Bull's ballistic experiments (1904) recorded at 1500 fps,
\url{<https://youtu.be/RLR-LT55Ueo>}.

\bibitem{Brixner:2017}
Webpage ``High-Speed Photography'' from the Atomic Heritage website,
\url{<https://www.atomicheritage.org/history/high-speed-photography>}.


\bibitem{Imaging-book}
Graham~Saxby,
\textit{The Science of Imaging. An Introduction}, 
2nd edition (CRC Press, 2010).

\bibitem{Applied-imaging-book}
Sidney~Ray,
\textit{Scientific Photography and Applied Imaging}, (Focal Press, 1999).

\bibitem{Fuller:2009}
P. W. W. Fuller,
``An introduction to high speed photography and photonics,''
The Imaging Science Journal \textbf{57} (6), 293--302 (2009).

\bibitem{Brandaris}
Chien Ting Chin, Charles Lancée, Jerome Borsboom, Frits Mastik, Martijn E. Frijlink, Nico de Jong, Michel Versluis, and Detlef Lohse,
``Brandaris 128: A digital 25 million frames per second camera with 128 highly sensitive frames,'' 
Rev. Sci. Instrum. \textbf{74} (12), 5026--5034 (2003).



\bibitem{Kubicki:2006}
Aleksander A. Kubicki, Piotr Bojarski, Marek Grinberg, Michał Sadownik, and Benedykt Kukliński, 
``Time-resolved streak camera system with solid state laser and optical parametric generator in different spectroscopic applications,'' 
Opt. Comm. \textbf{263} (2), 275--280 (2006).



\bibitem{Kaproth:2015}
Katherine A. Kaproth-Joslin, Refky Nicola, and Vikram S. Dogra, 
``The history of US: from bats and boats to the bedside and beyond,'' 
Radiographics \textbf{35} (3), 960--970 (2015).

\bibitem{OBrien:2015}
William D. O'Brien Jr and Floyd Dunn, ``An early history of high-intensity focused ultrasound,''
Phys. Today \textbf{68} (10), 40--45 (2015).

\bibitem{Ultrasonics-book}
Dale Ensminger, Leonard J. Bond,
\textit{Ultrasonics: Fundamentals, Technologies, and Applications}, 
3rd Edition (CRC Press, 2011).

\bibitem{Springer-acoustics}
F. Dunn, Thomas Rossing, W.M. Hartmann, D.M. Campbell, N.H. Fletcher (editors),
\textit{Springer Handbook of Acoustics},
2nd edition (Springer, 2014).



\bibitem{Rahman:1974}
Aneesur Rahman and Frank H. Stillinger,
``Propagation of sound in water. A molecular-dynamics study,''
Phys. Rev. A \textbf{10}, 368--378 (1974).

\bibitem{Teixeira:1985}
J. Teixeira, M. C. Bellissent-Funel, S. H. Chen, and B. Dorner,
``Observation of New Short-Wavelength Collective Excitations in Heavy Water by Coherent Inelastic Neutron Scattering,''
Phys. Rev. Lett. \textbf{54}, 2681--2683 (1985).

\bibitem{Ruocco:1999}
Giancarlo Ruocco and Francesco Sette, ``The high-frequency dynamics of liquid water'', 
J. Phys: Cond. Matter \textbf{11} (24), R259--R293 (1999).

\bibitem{Santucci:2006}
S. C. Santucci, D. Fioretto, L. Comez, A. Gessini, and C. Masciovecchio,
``Is There Any Fast Sound in Water?,''
Phys. Rev. Lett. \textbf{97}, 225701 (2006).


\bibitem{Zaleski:2015}
R.~Zaleski, 
``Principles of positron porosimetry,'' 
Nukleonika \textbf{60} (4), 795--800 (2015).

\bibitem{Gidley:2006}
David W. Gidley, Hua-Gen Peng, and Richard S. Vallery, 
``Positron annihilation as a method to characterize porous materials,'' 
Ann. Rev. Mat. Research \textbf{36} (1), 49--79 (2006).


\bibitem{Griffiths:2008zz}
David~J.~Griffiths,
\textit{Introduction to elementary particles},
2nd edition (Wiley-VCH, 2008).

\bibitem{Nahar:2020}
Sultana~N.~Nahar and Bobby~Antony, ``Positron Scattering from Atoms and Molecules,'' 
Atoms 2020, 8 (2), 29.

\bibitem{Cassidy:2007}
David B. Cassidy and A. P. Mills, 
``The production of molecular positronium,'' 
Nature \textbf{449} (7159), 195--197 (2007).


\bibitem{Shank:1983}
C. V. Shank, R. Yen, and Ch. Hirlimann, ``Time-resolved reflectivity measurements 
of femtosecond-optical-pulse-induced phase transitions in silicon,'' 
Phys. Rev. Lett. \textbf{50} (6), 454 (1983).

\bibitem{Rousse:2001}
Antoine Rousse, Christian Rischel, S. Fourmaux, Ingo Uschmann, Stéphane Sebban, G. Grillon, Ph Balcou et al. 
``Non-thermal melting in semiconductors measured at femtosecond resolution,'' Nature \textbf{410} (6824), 65--68 (2001).

\bibitem{Medvedev:2015}
Nikita Medvedev, Zheng Li, and Beata Ziaja, 
``Thermal and nonthermal melting of silicon under femtosecond x-ray irradiation,'' 
Phys. Rev. B \textbf{91} (5), 054113 (2015).


\bibitem{Workman:2022ynf}
R.~L.~Workman et al,
``Review of Particle Physics,''
PTEP \textbf{2022}, 083C01 (2022).


\bibitem{Weinberg-discovery}
Steven~Weinberg,
\textit{The Discovery of Subatomic Particles}, 
2nd edition (Cambridge University Press, 2003).


\bibitem{Marteau:2012zv}
J.~Marteau, D.~Gibert, N.~Lesparre, F.~Nicollin, P.~Noli and F.~Giacoppo,
``Muons tomography applied to geosciences and volcanology,''
Nucl. Instrum. Meth. A \textbf{695}, 23--28 (2012).

\bibitem{Morjean:2008}
M. Morjean \textit{et al,} 
``Fission time measurements: A new probe into superheavy element stability,'' 
Phys. Rev. Lett. \textbf{101} (7), 072701 (2008).

\bibitem{Busch:2008}
Paul Busch, ``The time–energy uncertainty relation,'' 
In \textit{Time in quantum mechanics}, pp.73--105, 
(Springer, 2008).

\bibitem{Duer:2022ehf}
M.~Duer \textit{et al.}
``Observation of a correlated free four-neutron system,''
Nature \textbf{606} (7915), 678--682 (2022).


\bibitem{Gibney:2022}
Elizabeth Gibney, ``How many yottabytes in a quettabyte? Extreme numbers get new names'',
Nature (2022), \url{<https://www.nature.com/articles/d41586-022-03747-9>}.

\end{thebibliography}
\end{document}